# Geometrical parameters dependence towards ultra-flat dispersion square-lattice PCF with selective liquid infiltration

Partha Sona Maji[*] and Partha Roy Chaudhuri

Department of Physics & Meteorology, Indian Institute of Technology Kharagpur-721 302, INDIA
*Tel: +91-3222-283842 Fax: +91-3222-255303,*
*Corresponding author: parthamaji@phy.iitkgp.ernet.in*

**Abstract:** We have performed a numerical analysis of the structural dependence of the PCF parameters towards ultra-flat dispersion in the C-band of communication wavelength. The technique is based on regular square-lattice PCF with all the air-hole of same uniform diameter and the effective size of the air-holes are modified with a selective infiltration of the air-holes with liquids. The dependence of the PCF structural parameters namely air-hole diameter and hole-to-hole distance along with the infiltrating liquid has been investigated in details. It is shown that the infiltrating liquid has critical influence on both the slope and value of dispersion, while pitch only changes the dispersion value whereas air-hole diameter modifies the slope of the dispersion. Our numerical investigation establishes dispersion values as small as 0±0.58ps/(nm-km) over a bandwidth of 622nm in the communication wavelength band (C-band). The proposed design study will be very helpful in high power applications like broadband smooth supercontinuum generation, ASE suppressed amplification etc.

## 1. Introduction

For communication systems, chromatic dispersion control in optical fibers is a very important problem, for both linear (like dispersion compensation etc) and nonlinear applications (especially for ultra short pulse propagation for broadband supercontinuum generation (SCG) etc). In all the cases, ultra-flattened dispersion fiber behavior becomes a crucial requirement.

Photonic Crystal Fiber (PCFs), also called holey fibers or microstructured optical fibers (MOFs) [1-2], possess the especially attractive property of great controllability in chromatic dispersion and nonlinearity by varying the air-hole diameter and hole-to-hole spacing. PCFs, because of their unique properties compared to bulk media and conventional fibers have been extensively investigated for realization of shifting ZDW [3-4], large negative dispersion [5-6] and ultra-flat dispersion in the required wavelength range [7-16]. Conventional PCFs use air-hole of same diameter arranged in a regular pattern throughout the cladding of the structure. However to realize ultra-flattened dispersion over a long wavelength window, we require different air-hole diameter in the cladding [11-16].

In this paper, we have presented that the effect of different air-hole diameter, without changing the air-hole diameter, can be realized with regular square-lattice PCF (S-PCF) geometry by infiltrating selective air-holes with various liquids such as polymers [17], water [18] and ethanol [19]. The concept of liquid infiltration in PCF has been found to be suitable for applications like Tunable PBG effect and long-period fiber grating [20]. In this work, a detailed analysis of the dependence of the structural parameters of S-PCF along with infiltrating liquid, upon total dispersion towards achieving ultra-flat near zero dispersion PCFs with selective liquid infiltration in the inner air-hole rings has been presented. The design establishes an ultra-flat dispersion as small as of $D$=0±0.58 ps/nm/km (with wavelength bandwidth of 622nm) over the communication wavelength band.

## 2. Design of Liquid filled PCF and analysis method:

Regular PCF consist of air-holes arranged in a regular pattern distributed symmetrically in the cladding. For conventional triangular-lattice PCF, we use *"d"* as the air-hole diameter and "Λ" as the hole to hole distance. In a square-lattice PCF, we use Λ as the distance between two air-holes in both horizontal and vertical directions with *d* as the air-hole diameter. S-PCF has been found to be a better candidate compared to its triangular counterpart in cases like wideband single-mode operation [21] and broadband dispersion compensation, higher effective area and red-shifting of Zero Dispersion Wavelength (*ZDW*) [4]. S-PCF is endlessly single-mode for higher *d*/Λ than triangular one [21] and it can better compensate the inline dispersion for broad dispersion compensation as it is having relatively closer relative dispersion slope (RDS) with existing SMF28 [4]. S-PCF provides higher effective area [4] that allows accumulating high power in the core region for high power applications like Supercontinuum Generation (SCG). Red-shifting of ZDW with S-PCF compared to triangular lattice PCF [4] allows broadband SCG in IR region especially with non silica materials. With these advantages we aimed to achieve near-zero ultra-flat dispersion nature with S-PCF. To design ultra-flat near zero dispersion PCF for wideband wavelength, we need variable air-hole diameter in the cladding [11-16]. The same can be achieved with uniform air-hole diameter by filling the air-holes with liquid of certain refractive indices and thereby mimic the effect of variable air-hole diameter. Depending on the refractive index (*RI*) of the infiltrating liquid, the effective size of the air-hole diameter can be modified.

There are certain issues related to the infiltration of liquid to the air-holes, whether the fluid wets glass and how viscous it is. If the liquid does not wet glass then surface tension will oppose entry of the liquid into the hole, making it difficult to fill. For this case we need a pressure greater than one atmosphere for air-hole diameter of values 0.46μm (our optimized diameter here). For this case a powerful vacuum pump will be required. On the other hand if the fluid does wet glass then the filling speed will depend on viscosity. Thus, we can fill the holes (and how quickly), with the given values for surface tension, contact angle and viscosity.

The schematic of the design has been presented in Fig. 1. The S-PCF consist of four air-hole rings with the inner air-hole ring infiltrated with liquid of refractive indices $n_L$. The inner air-hole ring can be infiltrated with certain liquid first by fusing the outer rings of air holes with fusion splicing technique [22] and then by immersing one end of the fiber in a liquid reservoir and applying vacuum to the other end of the fiber [18]. As the *RI* of the infiltrating liquid is less than the background silica, light will be guided by the modified total internal reflection (*TIR*). Modal field properties are evaluated by with CUDOS MOF Utilities [23] that uses Multipole method [24-25]. The efficiency and validity of the method has already been presented [24-25]. The total dispersion (*D*) is computed with

$$D = -\frac{\lambda}{c}\frac{d^2 \operatorname{Re}[n_{eff}]}{d\lambda^2} \quad (1).$$

Here $Re[n_{eff}]$ stands for the real part of the effective indices obtained from the simulations and *c* is the velocity of light in vacuum.

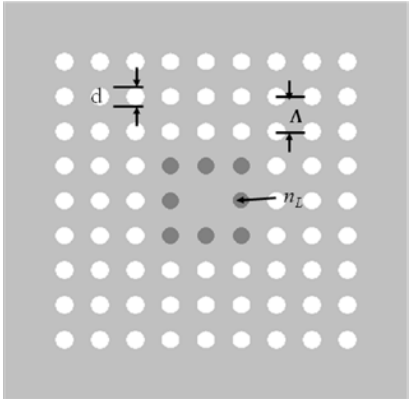

Fig 1: Cross section of the proposed photonic crystal fiber. The shaded regions represent air holes infiltrated with liquid with refractive indices $n_L$.

## 3. Structural parameters dependence towards ultra-flat dispersion

Our approach of optimization relies on varying multi-dimensional parameter space that consists of the liquid *RI* ($n_L$), hole-to-hole distance (Λ), and air-hole diameter (*d*) to design ultra flat, near zero dispersion optical fibers. The optimization process has been considered through few steps. In the first step, we'll study the dependence of the individual structural parameters along with the infiltrating liquid RI towards total dispersion. For this step, we'll consider a liquid with constant *RI*. In the second step, we'll optimize a design relying upon the findings of the first stage. In the third stage we'll choose a practical liquid (wavelength dependent) corresponding to the RI of the liquid optimized in the second stage. We'll re-optimize the other parameters with the available liquids to achieve near zero ultra-flat dispersion around the communication wavelength.

The first step of the optimization process is presented here, with Fig. 2 shows the effect of Λ on the *D* values. The figure reveals that the dispersion values changes for different values of Λ without much change in the slope. So, changing Λ has the effect of total dispersion. Figure 3 presents the effect of air-hole diameter (*d*) upon total dispersion. The dispersion slope changes rapidly for higher values of air-hole diameter where as the effect is less prominent for lower values of air-filling fraction (*d*/Λ). This can be attributing to the fact that with the increase of *d*, the waveguide effect increases and the interplay between material dispersion and waveguide dispersion results in the oscillation of the dispersion slope. Thus, changing *d* has the effect of changing dispersion slope. The effect of $n_L$ upon dispersion has been presented through Fig. 4. For smaller values of $n_L$ the slope is always positive where as the slope increases first and then decreases for higher wavelength for higher values of $n_L$. The figure clearly presents that, changing $n_L$ is having the effect of modifying both dispersion and its slope. Thus, we summaries the above effect as *varying the Λ influences the total dispersion while varying d has the desired effect of modifying the dispersion slope, and varying $n_L$ modifies both*.

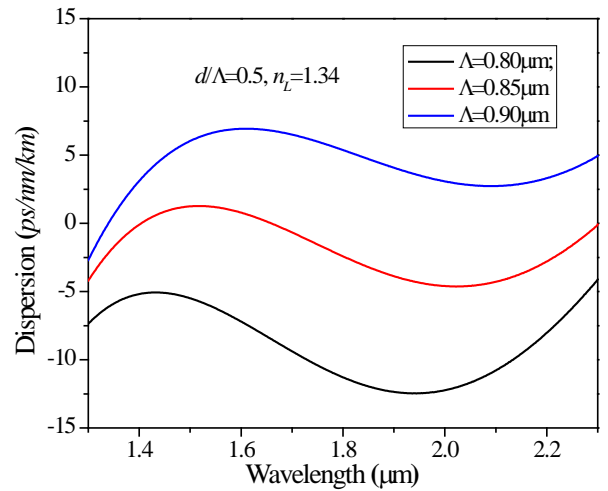

Fig 2: Dispersion variation of the PCF as a function of Λ keeping $n_L$ and *d* fixed

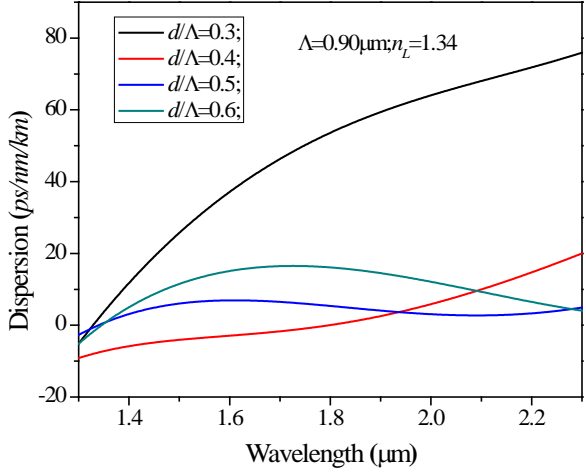

Fig. 3: Variation of Dispersion as a function of air-hole diameter (*d*) with Λ and $n_L$ remain constant.

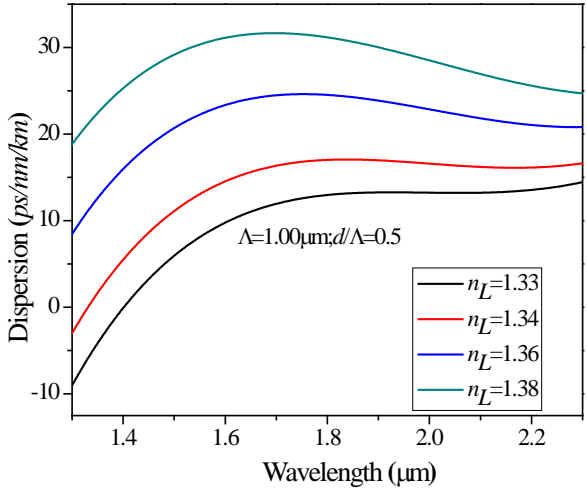

Fig. 4: Dispersion performance as calculated for varying $n_L$ values keeping pitch (Λ) and *d* fixed

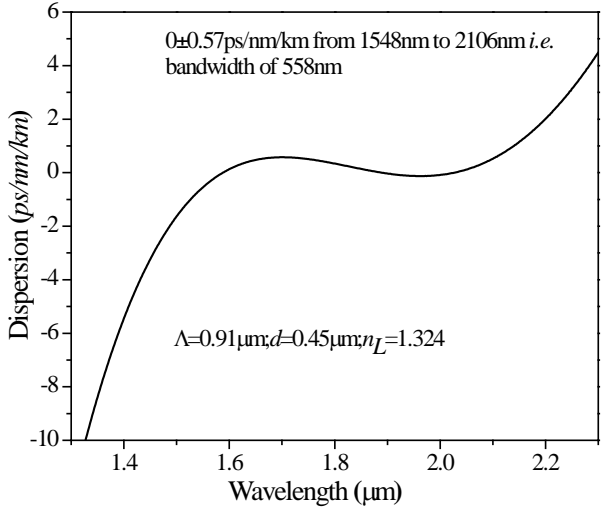

Fig. 5: Ultra-flat dispersion of 0±0.57ps/nm/km over 1548-2106nm for a bandwidth of 558nm for Λ=0.90μm, $n_L$=1.324, *d*=0.45μm.

## 4. Optimization towards near-zero ultra-flat dispersion PCF

With the help of the above conclusion regarding the dependence of the structural parameters upon dispersion, we move towards the second stage of our aim. We started with $n_L$=1.35 and Λ=1.00μm and we changed *d* to get an flat dispersion not necessarily near zero. Next, we optimize the other parameters keeping the previously obtained *d* fixed. After changing the parameters (either lowering or raising the values of the parameters) we have obtained one ultra-flattened near zero dispersion PCF as shown in Fig. 5. The figure reveals an ultra-flattened dispersion of D=0±0.57ps/nm/km from 1548nm to 2106nm *i.e.* for a wavelength bandwidth of 558nm. The optimized parameters for this purpose was found to be Λ=0.91μm, *d*=0.45μm and $n_L$=1.324.

Now having obtained preliminary design of near zero ultra-flattened dispersion with an artificial liquid we moved towards the final stage of our target of achieving near-zero ultra-flattened dispersion PCF with practical liquid. For this purpose we considered a liquid (calling liquid#1) available with M/S Cargille Lab. Inc. USA [26] whose *RI* is close to the previously optimized value. With this liquid we could achieve an ultra-flattened dispersion of D=0±0.76 ps/nm/km from the wavelength range from 1540nm to 2196nm *i.e.* for a bandwidth of 656nm with Λ=0.91μm and *d*=0.46μm as shown in Fig. 6. As the liquid *RI* is wavelength dependent, the material dispersion of the liquid is an important aspect and the same is presented in Fig. 7. The figure clearly indicates significant contribution of the liquid towards the target dispersion. The flexibility of the design procedure has been presented in Fig. 8 which presented another ultra-flattened dispersion with a different liquid (calling liquid#2) with a different set of optimized parameters. With liquid#2 we could achieve an ultra-flat dispersion of D=0±0.58 ps/nm/km for the wavelength range from 1664nm to 2286nm *i.e.* for a bandwidth of 622nm with Λ=0.93μm and *d*=0.50μm.

Cauchy equation of the oils:

Liquid#1: $n_1(\lambda)=1.3241434+201757/\lambda^2+5.656121\times10^{10}/\lambda^4$ (2)

Liquid#2: $n_2(\lambda)=1.3146073+184118/\lambda^2+1.095603\times10^{11}/\lambda^4$ (3)

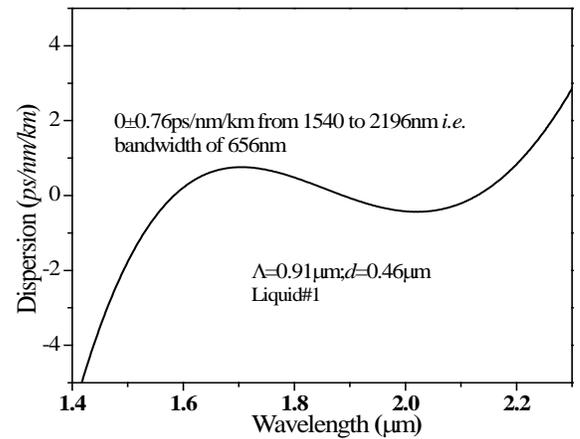

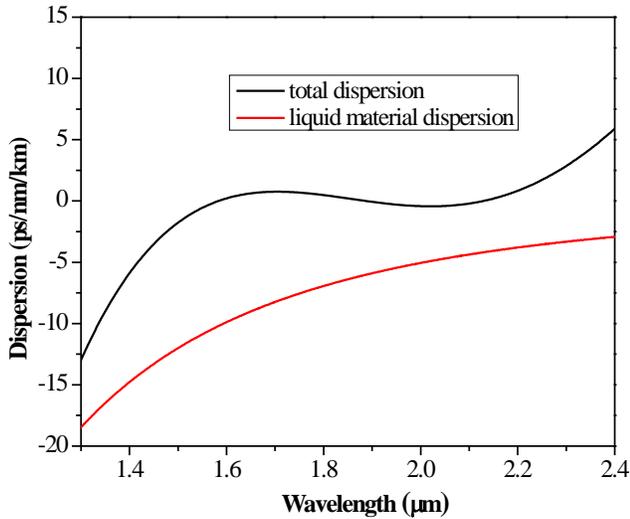

Fig 6: The ultra-flat dispersion of 0±0.76ps/nm/km over 1540-2196nm with a bandwidth of 656nm obtained with Liquid#1 with Λ=0.91μm and d=0.46μm.

Fig 7: Contribution of the material dispersion of Liquid#1 towards the total dispersion for the fiber with Λ=0.91μm and d=0.46μm. Material dispersion of the liquid contributes significantly towards achieving ultra-flat near zero dispersion.

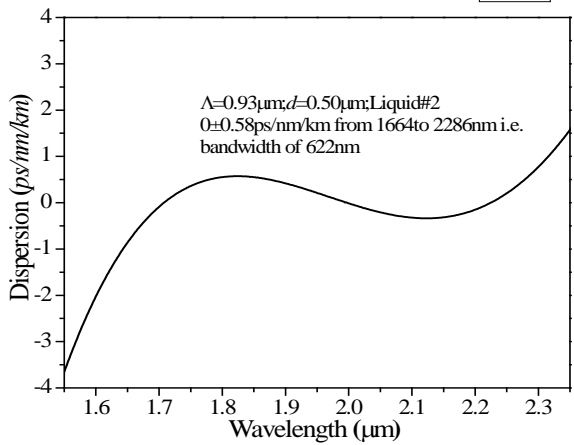

Fig 8: The ultra-flat dispersion of 0±0.58ps/nm/km over 1664-2286nm with a bandwidth of 622nm obtained with Liquid#2 with Λ=0.93μm and d=0.50μm.

## 5. Conclusions and discussions

Towards designing near zero ultra-flattened dispersion for a wide wavelength window we have performed a numerical analysis of the the structural parameter dependence of uniform air-hole square-lattice PCFs with the first air-hole ring infiltrated with liquid of certain refractive indices. The study reveals that *varying d has the desired effect of modifying the dispersion slope whereas varying the Λ influences the total dispersion, and varying $n_L$ modifies both*. With the help of the above study we could achieve a near zero ultra-flattened dispersion PCF with D as small as 0±0.58 ps/nm/km in for a bandwidth of more than 600nm. Two designs with two different available liquids have been optimized. Our design will have great influence on many engineering applications, namely dispersion compensation over wide wavelengths, birefringence control, wideband smooth supercontinuum generation, ultra-short soliton pulse propagation and many other photonic device applications like PBG devices and long period fiber gratings. Design and study of broadband smooth SCG spectra based on near-zero ultra-flat dispersion PCF is currently under study [27].

## Acknowledgements

The authors would like to thank Dr. Boris Kuhlmey, University of Sydney, Australia for providing valuable suggestions in understanding the software for designing and studying the properties of different structures. The authors acknowledge sincerely the Defence Research and Development Organization, Govt. of India and CRF of IIT Kharagpur for the financial support to carry out this research.